# Mixed Anhydrides at the Intersection Between Peptide and RNA Autocatalytic Sets: Evolution of Biological Coding

S.A. Kauffman[1,*] & N. Lehman[2]

[1]Institute for Systems Biology, Seattle, WA
[2]EDAC Research, P.O. Box 26276, Milwaukie, OR

*Author for correspondence: S.A. Kauffman (stukauffman@gmail.com)

## Abstract

We present a scenario for the origin of biological coding. In this context, coding is a semiotic relationship between chemical information stored in one location that links to chemical information stored in a separate location. Coding originated by the cooperative interaction of two, originally separate collectively autocatalytic sets, one for nucleic acids and one for peptides. When these two sets interacted, a series of RNA-folding-directed processes led to their joint cooperativity. The amino acyl adenylate, today amino acid-AMP, was the first covalent association made by these two collectively autocatalytic sets and solidified their interdependence. This molecule is a palimpsest of this era, and is a relic of the original semiotic, and thus coding, relationship between RNA and proteins. More defined coding was driven by selection pressure to eliminate waste in the collective autocatalytic sets. Eventually a 1:1 relationship between single amino acids and short RNA pieces (*e.g.*, three nucleotides) was established, leading to what is today known as the genetic code. Transfer RNA aminoacylating enzymes, or aaRSs, arose concomitantly with the advent of specific coding. The two classes of aaRS enzymes are remnants of the duality of complementary information in two nucleic acid strands, as originally postulated by Rodin and Ohno. Every stage in the evolution of coding was driven by the downward selection on the components of a system to satisfy the Kantian whole. Coding was ultimately forced because there were at least two chemically distinct classes of polymers needed for open-ended evolution; systems with only one polymer cannot exhibit this characteristic. Coding is thus synonymous with life as we know it, and can be thought of as a phase transition in the history of the universe.

**Keywords:** Origins of life, coding, RNA, peptides, autocatalytic sets, mixed anhydrides

1. **Introduction**

The evolution of the genetic code is a massive mystery; there have been thousands of papers theorizing its origin and evolution. But it is important to realize that the origins of biological coding as a phenomenon and the origins of the genetic code *per se* were fundamentally different events. The former preceded the latter, and it is possible that the two events were causally related.

The relationship between coding and the genetic code has been queried, mainly from the viewpoint of information theory. Gatlin [1] and later Yockey [2] focused on the parallels between computer coding and the biological genetic code. These authors emphasized the mapping aspect of the genetic code, and the types of functions that could map a set of nucleotides onto a set of amino acids. Yockey [2] framed the problem in terms of set theory and reiterated a definition of a code from that perspective as, "a unique mapping of the letters of alphabet *A* on to the letters of alphabet *B*," following Perlwitz *et al*. [3].

No matter how it is perceived, coding is a situation in which information about an object or event is stored somewhere else. This is, at heart, an issue of semiotics. Because biology is a subset of chemistry having unique processes, we therefore need to consider semiotics from a chemical point of view. One informational chemical must point to another, as a sign, a signal, or an icon. In the realm of nascent life, we must figure out why, and how, coding arose, and then deduce its downstream influences on the biotic world. Davies noted that "real" life required coding; you can envisage trivial replicators but until you have coding, they do not advance [4].

Today, there is a distinct co-linearity between information stored in nucleotides (RNA) and information stored in amino acids (polypeptides). We perceive the information thus as "coded" such that there must be a decoding apparatus that is responsible for chemically reconfiguring (translating) the information stored in RNA into polypeptides. This decoding is carried out by the joint action of aminoacyl tRNA (aaRS) enzymes, tRNAs, and the ribosome. Humans can refer to the "genetic code" as a look-up table to see the one-to-one mapping of RNA information in codon (or anticodon) triplets and the amino acids that they specify. The question of why certain codons specify their cognate amino acids is a long-standing one in biology and has been the subject of 60 years of (mostly theoretical) inquiry. But the more fundamental question of why a code exists in the first place, and how it came to be—chemically and evolutionarily—remains open and understudied in origins research. The simple answer that information storage is best in nucleic acids and information manifestation is best in proteins, and therefore there *must* be some code, is not sufficient. At first glance the chemistry of nucleotides and the chemistry of amino acids are so different that it is not apparent how a code could have arisen *de novo*. We need a theory of coding origins that takes into account information theoretic, chemical, and evolutionary considerations to reconstruct a reasonable history of how RNA/peptide co-linearity gained a foothold in biology.

The most thorough and thoughtful analysis of the origins of coding have been performed by Carter & Wills [5–8]. Their work focused on the origin, evolution, and contemporary (and historical) substrate specificity of aaRS enzymes. A major takeaway from their analyses is that coding did not originate with catalytic RNAs (ribozymes), but instead was first established in archaic polypeptides, a conclusion reached by considering the reflexivity possible in aaRS and protein-synthesis functions, cf. [5]. They derive insight from the observation that there are two distinct classes of aaRS, and that one can deduce an ancestral "gene" on which these two classes are encoded on opposite, and thus interdependent, nucleic-acid strands [9,10]. Class I enzymes, which tend to deal with the larger amino acids, have a catalytic core (HIGH/KMSKS) that involves amino acids that must be charged by class II enzymes (*i.e.*, H, G, K, and S). The converse is true, suggesting an "ancient hypercycle-like interdependence" of the two enzyme classes [5]. This leads to a model of feedback loops that argue for a self-supporting (autocatalytic) protein world but not a self-supporting ribozymal world. Importantly, Carter & Wills downplay the primordial role of the ribosome in coding origins, and instead posit that coding predated the earliest encoded peptides, or at least the two events were contemporaneous [6,7].

An alternative view can be derived from a careful analysis of the ribosome, and this has led to an inquiry into the relationship between the peptidyl-transferase center (PTC) and the coding phenomenon [11–14]. This view is that in the milieu of the proto-ribosome a symbiotic relationship between proteins and nucleic acids was birthed [14]. The proto-ribosome would have been a loose collection of RNA stem-loops and divalent cations (first $Fe^{2+}$, later $Mg^{2+}$) having rudimentary catalytic activity. Although the origins of the original aaRS activity are not explicit in this model, one can infer that either aaRS specificity was broad at the root of the ribosomal tree, and/or that the elements of the ribosome itself originally provided a rudimentary mapping of amino acids onto nucleotide sequences [14,15].

Both the aaRS-centric view and the ribosome-centric view provide valuable insights into the history of protein synthesis and frame the macromolecular events that surrounded the advent of the coding phenomenon. However, neither adequately explains how coding itself came to be in the first place, nor what the ultimate chemical innovation was that cleared the path for the origins of life. In this paper we propose that the mixed anhydride bond was just this innovation, and that its inclusion into autocatalytic sets was the spark of life.

## 2. Results

2.1. Chemical semiotics

To understand how one chemical acts as a sign (*etc*.) for another, we must consider how the earliest group interactions played out during the origins of life. Carter & Wills begin to address this semiotic issue. They state that, "The earliest genetic coding paradigm therefore required simultaneously solving three different recognition problems—ATP, amino acid, and tRNA—and finding two related catalytic mechanisms, each in two different ways (for Class I and Class II aaRS) in order to implement all events necessary to accomplish the symbolic conversion" [7]. But the question remains, where do the aaRS (and their embodiment in a double-stranded RNA = dsRNA) come from?

We have argued, implicitly, that the answer to this lies in a more fundamental—and pervasive—phenomenon, that of the collective autocatalytic set (CAS) [16–19]. A CAS is a collection of molecules that spontaneously forms a network of interdependent catalytic connections to ensure self-propagation of the whole set. We would describe the aaRS as an entity that a resulted from the intersection between two (previously independent) CASs: the RNA CAS and the peptide CAS. This intersection would be the first creation of a covalent bond between a component of the first (*i.e.*, a nucleotide) and a component of the second (*i.e.*, an amino acid). We represent this covalency as a black dot in Figure 1. Such a hybrid molecule must originally have had some role in benefiting *both* CASs, and its subsequent transfer to a tRNA must have been a much later invention/requirement.

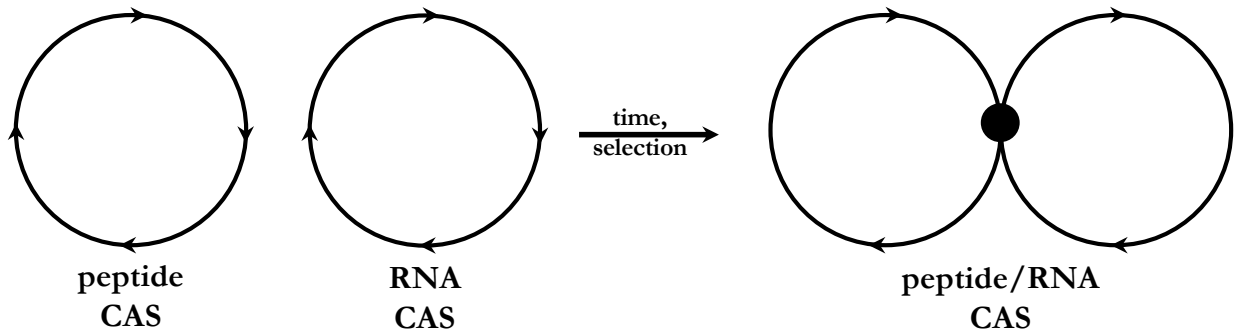

**Figure 1**. The merging of two separate autocatalytic sets (CAS) to form a more complex one in which coding exists.

2.2. The mixed anhydride model of the origins of life

There is one obvious molecule that fits the description above perfectly: the aminoacyl adenylates. These are single amino acids that have been "energized" with the formation of a covalent bond to an adenosine nucleotide. In reactions catalyzed today by the aaRS, an ATP molecule is bound to an amino acid forming a mixed anhydride bond: a physical link between a nucleic-acid precursor and a polypeptide precursor (Figure 2). Once formed, they serve as the building blocks for ribosomal-directed protein synthesis.

One problem to solve here is, why ATP (as compared to GTP, UTP, or CTP)? The answer to this question would address a long-standing enigma in biology. It could be as simple as the aqueous concentration of adenine was higher than any other nucleobase. Adenine was shown decades ago by Oró to be produced in roughly 1% yield as hydrogen cyanide (thought to be prebiotically abundant on the Earth) is heated at 60°C [20]. Or it could have been that the formation of the adenosine nucleotide (*e.g.*, AMP, ADP, ATP, or AppA) is/was thermodynamically more accessible than that of any other nucleotide.

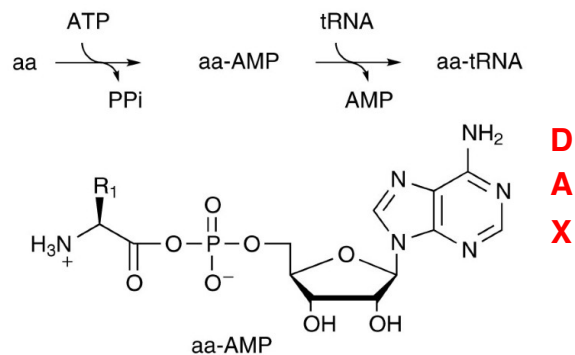

**Figure 2.** The formation of an aminoacyl AMP via the creation of a mixed anhydride bond. These two reactions are performed today by the action of aaRS enzymes, whose substrate specificity manifests the genetic code.

Another possible explanation for the primacy of ATP and its utilization in energy transfer, including amino-acid charging is that the semiotic nature of the Watson-Crick surface of ATP was either the simplest of them all, or the easiest from which expansion into a larger code could take place. Reading from the most distal moiety from the glycosidic bond down the W-C face of a nucleobase, adenine reads (hydrogen bond) donor-acceptor-none (DAX; red letters in Fig. 2). Uracil would read ADX, being complementary to adenine, in the canonical pairing orientation. For completeness, guanine would read ADD and cytidine would read DAA.

From a binary coding perspective, the simplest possible way that four states can be encoded elsewhere is through two positions of a 0/1 code (Code I in Table 1). However, if one allows for additional positions for either redundancy or punctuation, one can have three or more positions (*e.g.*, Codes II and III in Table 1). There is an analogy between Code I below and purine nucleobase pairing and Code II below and pyrimidine nucleobase pairing. If the original coding only involved purines, then it could follow Code I: DAX–ADX. But pyrimidines would require Code II (DAD–ADA, DDA–AAD, *etc.*). Clearly then, a proto-biological system based on adenosine (and a complement, possibly uracil) would require the least complexity.

Table 1. Three Possible Binary Codes for Four-Letter Alphabet

| Letter | Code I | Code II | Code III |
|--------|--------|---------|----------|
| A      | 0 0    | 0 0 0   | 0 0 0 0 0 |
| B      | 0 1    | 0 1 1   | 0 0 1 1 1 |
| C      | 1 0    | 1 0 1   | 1 1 0 0 1 |
| D      | 1 1    | 1 1 0   | 1 1 1 1 0 |

Note that in the mixed anhydride, the P–O–C bond is formed very far from the W-C surface, especially when the nucleotide is in the *anti* configuration. Consequently, the information is linked from one physical location (the W-C surface), through the ribose foundation, to another, the distal end of the molecule where the amino acid identity ($R_1$) lies. The W-C surface thus acts as a sign (*sensu* Peirce) to point to a relatively distant object. Though today all mixed anhydrides used by aaRS enzymes are adenylates, at the time of the first intersection between RNA and peptide CASs, there may have been many covalent associations between nucleotides and amino acids. Eventually the sign became covalently disambiguated with the object, giving rise to a code (see below) and only aminoacyl-AMP remains as a palimpsest of that era.

## 2.3. A scenario of how RNA and proteins first interacted in a coding fashion

Two contemporaneous CASs, one of nucleotides/RNA, and one of amino acids/peptides, could have existed, and their cooperation could have benefited both [17]. Laboratory research has shown that RNAs alone [21] or peptides alone [22] can form self-reproducing networks. As such, they have the capacity to attain catalytic, constraint, and task closure, and do thermodynamic work to construct themselves [19,23]. However, the growth potential of these is limited; only sub-exponential reproduction can be achieved. Without exponential reproduction, molecular networks cannot escape competing selfish parasites, and cannot evolve complexity [24–27].

Thus, the critical mutual benefit that each CAS would offer each other is the ability to achieve exponential reproduction. From a nucleic acid standpoint, the barrier to sustained reproduction is strand melting. This problem manifests itself as the "strand displacement problem" and has been a long-standing obstacle to the synthetic creation of RNA autoreplicase ribozymes [28–30]. It is also the reason why its solution in the context of the PCR reaction revolutionized biological study and clinical practice: thermostable protein enzymes can catalyze strand separation and allow for exponential reproduction (replication in this case). From a protein standpoint, the opposite problem, of sorts, exists. Polypeptides do not have a reliable pattern of hybridization; one sequence cannot template another with a high degree of certainty and under a wide variety of environmental conditions. An enormous and phase transitionary advantage would result if certain amino acids or short peptides could have facilitated the melting of double-stranded RNAs, while certain nucleotides or proto-anticodons could have facilitated the annealing and/or ligation of amino acids.

A scenario of mutualism therefore presents itself (Figure 3). Originally there are two autocatalytic sets, one for peptides and one for RNAs. They operate independently, having been spawned from a prebiotic soup of small organics that included amino acids and nucleotides. One feature of these autocatalytic sets in our scenario is that they both run on recombination reactions primarily: *trans*-peptidation in the case of the peptide CAS and *trans*-esterification in the case of the RNA CAS. Other models of polymer CAS have been based mainly on cleavage/ligation reactions, and we do not preclude this mechanism although it requires a higher degree of chemical activation and is thus deemed less likely.

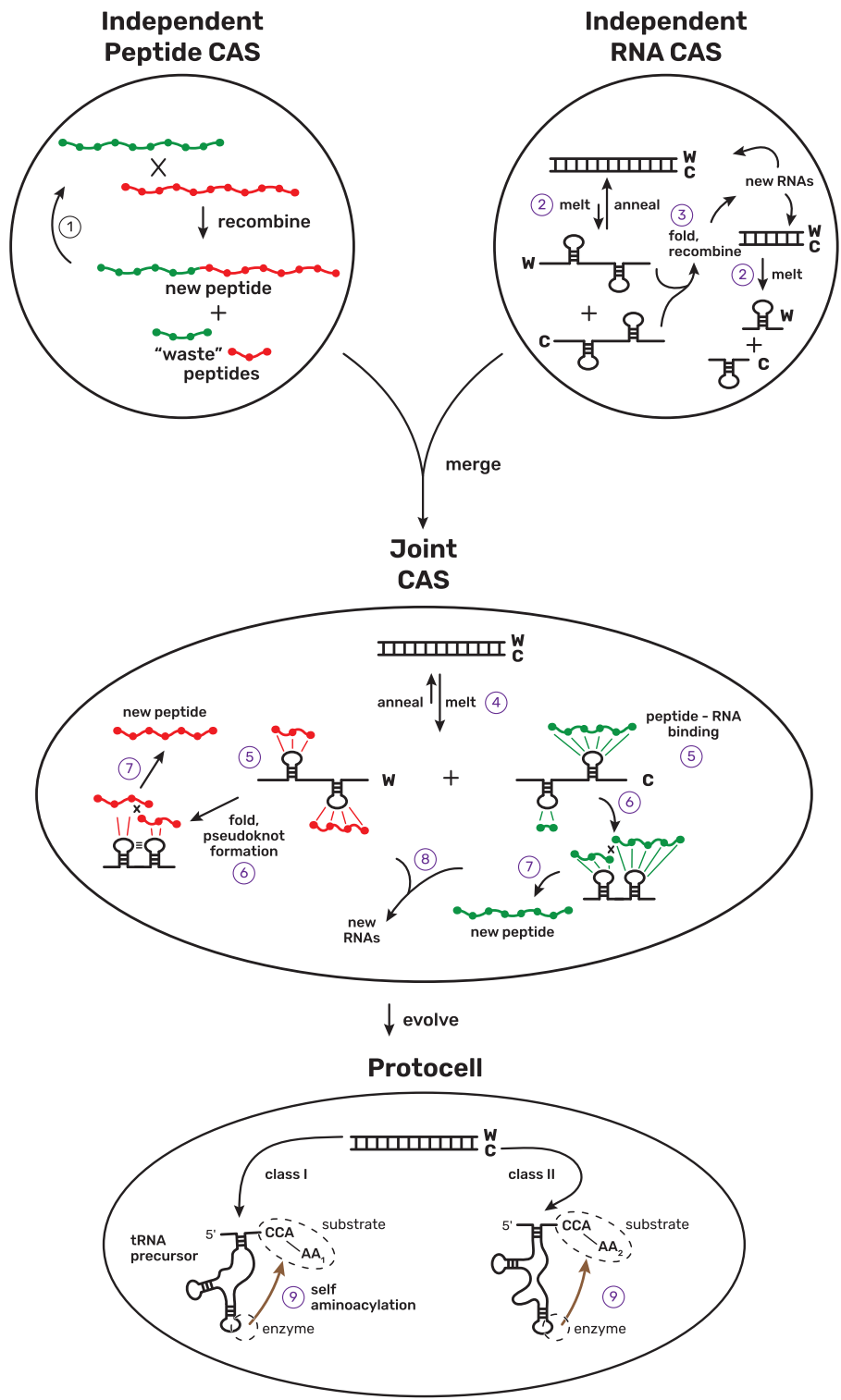

**Figure 3.** Scenario for the origins of coding. Originally there are two independent collectively autocatalytic sets (CASs), a peptide CAS and an RNA CAS. In each, reproduction is possible on its own. In the peptide CAS, two

peptides, green and red, if of sufficient length have the capacity to recombine to make new peptides. Small peptides of length, say, five amino acid or less, do not possess sufficient complexity to recombine (or ligate, following [22]). They are "waste" peptides, having low affinity for other short peptides. ① Longer products of reactions interact with other long products to create reaction cycles. Similar processes occur in the RNA CAS [21]. ② However, in the RNA CAS there are two complementary strands of RNA upon annealing, a "Watson" strand (**W**) and a "Crick" strand (**C**). Annealing is favored when two such strands interact; double-stranded RNA also helps protect the RNA information from spontaneous hydrolysis. ③ If RNAs do melt, longer strands can adopt secondary structures including stem-loops that engender catalytic activity. Shorter RNAs melt more easily, but have less tendency to possess catalytic activity. In both the independent CASs, reproduction is slow, sub-exponential, because of the structural, kinetic, and thermodynamic barriers to proper catalytic events. Nevertheless, each CAS can operate without regard for the other. If, however, the two CASs find themselves in a situation in which they can interact productively, then the stage is set for coding, and life. In a joint CAS, each polymer provides assistance to the other, enhancing the other's reproductive rate. ④ Interaction with peptides shifts the equilibrium of the dsRNA to the melted, and potentially catalytically active, forms: ⑤ peptides, long or short, can bind to the loop regions of melted RNA strands. ⑥ The folding of single-stranded RNA into more complex stem-loops and pseudoknots, helps bring the peptides, including the shorter ones, into close contact so that ⑦ they can catalytically recombine/ligate to form larger peptides. One subset of peptides develops a tendency to bind the **W** strand of RNA, while another subset develops a tendency to bind the **C** strand. The advent of exactly two distinct subsets is the result of there being exactly two strands of RNA that form double-stranded RNA [10]; dsRNA is the most thermodynamically stable nucleic acid complex. ⑧ Recombination/ligation of the RNAs proceeds as in the independent CAS case, although now its rate is augmented, eventually (though group selection) becoming exponential. Likewise, peptide reproduction is augmented because of the ability of smaller peptides to participate in the catalytic cycle. In the bottom, protocell stage, evolution drives the formation of the precursors to modern-day RNAs and proteins. Shorter and shorter peptides are selected for RNA loop binding, as the march to a 1:1 specificity between amino acids and nucleotide (triplets) is favored by further reduction of waste (see text). ⑨ The RNAs are selected to covalently attach the single amino acids to their ends, forming the aminoacyl adenylates that are now palimpsests of this era. Coding arises in full when a unique associate between specific amino acids and specific RNA sequences becomes established; life is a consequence of the duality of polymer types and their association through this coding. Self-aminoacylation is the first aaRS function, shared with tRNA function in this model; the set of reproducing RNAs includes the precursors to tRNA, rRNA, and aaRS activities, which later diverge and/or become taken over by peptides. The **W** and **C** strands of RNA each drive the evolution of the class I and II aaRS enzymes [10]. At each stage in this scenario, selection for the reproduction of the whole drives the evolutionary relationships of the parts (*i.e.*, a Kantian whole).

Suppose that the "Watson" (**W**) strand and the "Crick" (**C**) strand of a dsRNA can each form one stem loop, but only when denatured from each other. There may be two, three, or many more such RNAs in a CAS, and a reproduction cycle is maintained *via* recombination (or perhaps template-directed replication). Yet because of slow melting, this reproduction is sub-exponential. The binding of an amino acid to the nascent loop of the stem-loops would shift the equilibrium from dsRNA (**W-C**) to separate stem-loops (**W** and **C**), each with an amino acid bound to the nucleotides in the loop region. This would be the equivalent of the modern-day anticodon, although at this point it would not yet be functioning in the same role. Nevertheless, by pure physical-chemical properties, each short contiguous set of nucleotides (*e.g.*, three, although this need not be so, nor a fixed value) would bind a "cognate" amino acid with some degree of specificity. This hypothesis, that trinucleotides bound amino acids prior to the full development of the (genetic) code has often been proposed in various forms (*e.g.* [31]), but most noticeably as the stereochemical theory. Yarus *et al*. [32] has demonstrated that there is a statistically significant correlation between the binding constants in solution for amino acids and their cognate anticodon triplets. Rodin *et al*. [33] similarly have shown that amino acids bind preferentially to their modern-day anticodons, at least to the second and third positions. In our scenario, the **W** or **C** strands, when dissociated from each other, bind single amino acids or short peptides.

At first this binding is rather weak and non-specific. These short peptides are too short to participate actively in the CAS; only peptides of length, say, eight amino acids or more, possess enough structural complexity to be *trans*-peptidation catalysts. The smaller peptides are essentially non-productive members of their CAS; they are akin to waste products. Yet upon binding to the loops in an RNA, they become positioned, and ordered, in a way such that their recombination or ligation to make longer oligomers becomes enhanced. At the same time, the binding not only helps to melt the dsRNA but also to stabilize the secondary structures of the nascent stem loops [34].

There is ample evidence that non-covalent binding between amino acids and RNA was an ancient event and one of important regulatory function [34]. Yarus [35] detected a specific and reversible binding site in the catalytic core of the *Tetrahymena* ribozyme for L-arginine. This amino acid, much more so than any of the other biological 19, or even the D-stereoisomer, binds to a triplet sequence of nucleotides at the guanosine binding site of the ribozyme that is critically involved in is catalytic (self-splicing) function, and is actually a competitive inhibitor [36]. Since this discovery, hundreds of specific interactions between short (*e.g.*, tri-) nucleotide sequences and amino acids have been revealed, such as those in natural riboswitches and artificial aptamers.

### 2.3.1. Benefits to the RNA

The strand separation would benefit the RNA CAS. When the **W** and **C** strands are apart, they are free to form more complex secondary structures, interacting with other regions of the same strand as well as distant regions of the opposite strand. This would facilitate the formation of catalytic structures, which, in RNA, are often made possible by pseudoknotted configurations. A pseudoknot is a specific non-symmetrical pattern of nucleotide pairing which allows for catalytic nucleotides to be positioned at the active site, and/or by placing strain on particular phosphodiester bonds. These configurations are seen in ligase, replicase, and HDV ribozymes, for example.

With conformational freedom and the catalytic capabilities that ensue, an RNA population would be more able to access the numbers and types of catalytic events (*e.g.*, recombination reactions) that would permit exponential growth [17].

### 2.3.2. Benefits to the peptides

The binding to stem loops would benefit the peptide CAS. As free peptides, their ability to associate with each other in orientations that are productive for (*trans*-peptidation/ligation) is limited. In the well-studied case of template-directed ligation in peptide networks, the reproduction efficiency depends not only on the kinetic order of the reaction, but also on the ratio of reactions that are template-assisted to those that are template-free [37]. Peptide ligation reactions that contain a certain degree of templating can self-organize into small cross-catalytic networks, while those that do not can only form random, disorganized collections with a low rate of reproduction. Binding to an RNA scaffold affords peptides the ability to utilize templates that are not part of their own network.

Two amino acids or peptides that are bound to the loop regions of two RNA stem loops can be positioned for recombination or ligation. One way to envision this is that one RNA strand, say the **W** strand, once free of its complementary **C** strand, forms two stem loops, which then position near each other in space through tertiary interactions. This type of conformation is seen in many ribozymes, such as the hammerhead [38]. Upon joining to form longer peptides, these molecules can better participate in the reactions of the peptide CAS. Short fragments are no longer waste; they become incorporated into the self-reproduction network.

2.3.3. Strand specificity sets the stage for coding

If two loops from the same strand of RNA form and bring their bound amino acids/peptides together to allow for recombination/ligation, a polarity develops that leads to a coding situation. Amino acids that bind to the loops formed from the **W** strand will preferentially be joined to one another, while amino acids that bind to the loops formed from the **C** strand. The reason for this is that loops on the same strand can interact through *intra*molecular rearrangement (*i.e.*, folding) far more readily than they can with loops on (now) dissociated strands through *inter*molecular interactions. RNA folding drives peptide elongation.

Returning to the model presented in Figure 3, there exists empirical support for the folding processes that bring the two amino acids (or short peptides) together upon rearrangement such that they can be recombined to form longer peptides. Consider that it has been shown that one ribozyme can have two distinct folds, each with its own unique catalytic activity. A fold with ligase activity or a fold with HDV self-cleavage activity are both accessible by single RNA sequences of length 90 nucleotide [39]. Moreover, HDV genomic/antigenomic sequences can be templates for each other's replication [40].

The intramolecular reinforcement leads to two subsets of amino acids, those affiliated with the **W** strand, and those affiliated with the **C** strand. Because, by definition, the **W** and **C** strands are complementary, the minimal energy state for the entire system would be a symmetric one in which half of the participating amino acids at least transiently associate in one sub-network, while the other half associate in another sub-network. It is clear that this situation presages that of the aaRS enzymes, for which today there are 10 in each of two classes, and is in agreement with Carter & Wills' observation of reciprocity between two 10-member aaRS collections [5].

At this point there would emerge a strong selection pressure for class uniformity and specificity. That is, peptides composed of pure class I (say) amino acids would become associated with the **W** (say) RNA strand, while peptides composed of class II amino acids would become associated with the **C** strand. This would not yet be coding *per se*, merely the advent of two clouds of informational-rich polymers.

Selection would strengthen their self-reinforcement. Let us explain why. If there are, say, 20 amino acids involved in the peptide CAS, then there are 8000 possible tripeptides. Let the set of *N* **W** stem-loops bind "overlapping" subsets of the 8000 tripeptides that have higher affinity for the **W** stem-loops. Similarly, let this set of *N* **C** stem-loops bind overlapping subsets of the 8000 tripeptides having higher affinity for **C** stem-loops. Now consider the manner in which the peptides help the RNA CAS: to bind the **W** stem-loops on the **W** strand, and bind the **C** stem-loops on the **C** strand and help melt the dsRNA so that they can reproduce in the RNA CASs. Consider two extreme cases:

> *Case 1:* Let there be a long pure **W** strand and its complementary long pure **C** strand. Let each strand have a modest number of stem-loops, say 4 to 11. The *N* **W** stems bind **W** peptides (*e.g.*, class I) and recombine or ligate them to create longer pure **W** polypeptides. The *N* **C** stem-loops bind **C** peptides (*e.g.*, class II) and recombine or ligate them to create longer pure **C** polypeptides. This is self-consistent: long **W** RNA strands have denatured to create **W** stem-loops, the complementary long **C** RNA strands have denatured to create **C** stem-loops. When these longer complementary RNA strands replicate, the **W** strand and the **C** strand can be pulled apart (melted) by binding, respectively **W** polypeptides and **C** polypeptides, and can reproduce exponentially.

> *Case 2:* By contrast, consider that the two complementary long RNA strands each have both **W** stem-loops and **C** stem-loops in more or less random order along the RNA strands. These stem-loops create polypeptides that are likewise random sequences of **W** and **C** amino acids. When the longer **W**-**C** and **C**-**W** RNA strands reproduce and **W**-**C** polypeptides try to bind **W** and bind **C** stem-loops they will, but in general be out of sequential register. Because of this, the **W**-**C** polypeptides will not bind the two **W**-**C** strands efficiently, so will not effectively melt the two strands apart.

Thus, *pure* **W** and *pure* **C** RNA strands and *pure* **W** polypeptides and *pure* **C** polypeptides will form sets that reproduce more effectively. This sets up the scenario that further engrains coding. There will be a selection pressure toward longer RNA complementary strands, one pure **W** and one pure **C**. The pure **W** strands will interact with pure **W** peptides on nearby regions of the strand that are become proximal as a consequence of folding. The **W** strands drive the recombination of these peptides to elongate them. The analogous processes are occurring on the **C** strands with **C** peptides.

2.3.3. True coding arises upon a covalent interaction between RNA and peptides

At this point in our scenario so far, the stem loops of RNA are not conveying any information about the peptides to any other location. There is simply a selected tendency for **W** stem-loops to interact with **W** peptides, and for **C** stem-loops to interact with **C** peptides. The contemporary tRNA-aminoacyl synthetases activities are not yet fully integrated into nascent life.

However, the key event that, in one stroke, led to coding and the origins of life was the formation of a covalent bond between the amino acid and the RNA strand to which it is bound. This could only occur at a free end of the RNA. And although the chemical states of the two ends of the RNAs would have been quite variable in a chemical soup, containing a wide mixture of alkylation (methylation), amidation, hydroxylation, and phosphorylation states, the last two such termini would be the most activated and thus amenable for bond formation. These would have been numerous to allow for RNA recombination/ligation anyway.

Again, the extant evidence for a mechanism of amino acid-RNA covalent attachment is plentiful. Turk *et al*. [41] demonstrated that a host of small and simple RNA motifs have the catalytic capacity to transfer an amino acid and covalently attach it to their 2′ hydroxyl groups on their 3′-terminal nucleotide (usually a uridine). In the most extreme case, the enzyme portion of the RNA could be as short as 5 nucleotides in length (Figure 4). The substrate uridine was found to exist most often on the 3′ end of the sequence CCU, which mimics the CCA terminus of contemporary tRNAs (the attachment site of amino acids).

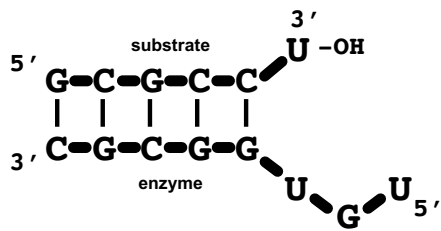

Figure 4. An example of a small self-aminoacylating ribozyme discovered by Turk *et al.* [41]. The top strand is the substrate, while the bottom strand is the enzyme (8 nt in this case).

The amino acid source for transfer for these mini-*trans*-aminoacylation ribozymes is already "activated" as an aminoacyl-AMP such as PheAMP (for phenylalanine) or MetAMP (for methionine). Therefore, the mixed anhydride bond must be formed prior to the activity of these mini ribozymes. Nevertheless, their existence and versatility portend two things. First, they suggest that RNAs could have performed at least one of the contemporary functions of aaRS enzymes, that of the charging of tRNA precursors. And second, they hint that the more rudimentary formation of the mixed anhydride bond, admittedly a greater thermodynamic challenge than aminoacylation, could have been a part of the small molecule CAS milieu prior to coding and life.

A progressive enhancement of the informational relationship between RNA sequences and amino acid identity solidified the coding. When, say, a stretch of **W** nucleotides was correlated with a five-amino-acid peptide, then the RNA is informationally linked to $n^5$ possible peptides. If *n* were twenty, and in contemporary biology, then this would mean $3.2 \times 10^6$ possible combinations could be associated with a particular RNA sequence, and the vast majority of these would be non-functional for the peptide-RNA collectively autocatalytic system. This in turn would lead to a tremendous amount of waste in the peptide CAS. Evolutionary pressure would then be strong to tighten the relationship, by shortening the peptide length. If the stretch of **W** nucleotides were correlated to only tripeptides, then the associated products would number about 8000, an efficiency improvement (in terms of waste in the peptide CAS) of many orders of magnitude.

As the length of peptides, *n*, progressively shortens from 5 to 4 to 3 to 2 to 1, the waste decreases. On the other hand, as the length of peptides drops, the chance that those peptides synthesized can also play a role in the peptide CAS decreases. Thus, as the length *n* decreases towards 1, reproduction of the system must shift from the peptide CAS to reproduction by *encoded* peptides. In sum, there is strong selection pressure to minimize waste, and the highest informational efficiency, would be achieved if there were a 1:1 correspondence between an RNA stretch and a single amino acid. At this point there would be true coding, and a covalent bond between RNA and a single peptide, as in the aminoacyl adenylate, would manifest this code. The length of the RNA codon stretch today is known to be three, as first discussed by Gamow in that three nucleotides is the minimum binary number capable of encoding a set of amino acids that exceeds 16 (*i.e.*, 20 or so).

Ultimately, each **W** and **C** RNA strands could evolve into the Rodin & Ohno duplex [10] as envisaged by Carter & Wills [5], and accordingly, **W**-encoded and **C**-encoded peptides could evolve into the Ur-aaRS enzymes as envisaged by Carter [42]. Of course, at some point during the honing of the peptide-RNA relationship towards 1:1, there was a transition to template-directed replication, both for peptides and for RNA. This contemporary form of reproduction post-dated coding. As hinted at by Carter & Wills though not fully articulated, symbolic coding eventually emulated hydrogen bonding [5], but the first key step was the formation of a covalent bond, now seen only in the mixed anhydride.

## 3. Discussion

We have presented a model of early polymeric molecular evolution that includes the origin and early development of coding. The relic of coding is the covalent bond found in the mixed anhydride molecule, central to contemporary translation. Coding was a direct consequence of the fact that two, chemically distinct polymers were needed for the genotype-phenotype duality that allows for full evolutionary freedom to explore fitness landscapes. An implication of this realization is that life as we know it could not have been possible without a code; simpler pre-life systems can undergo change but cannot encapsulate the characteristics of their environment into their "genomes," and thus tend to end up in uninteresting and/or closed-ended patterns such as stable limit cycles or parasitic dualities (cf. [4]). In fact, we can make the claim that coding *is* life, or at least that coding is a necessary condition for life. As stated eloquently by Davies: "life = matter + information." We have added coding: life = matter + information + coding [4].

Coding arises from the existence of two distinct biopolymers. We have discussed polypeptides and polynucleotides. However, we do not want to rule out the possibility that other polymers, particularly lipids, played key informational roles in the instigation of coding. Damer and Deamer for example, proposed an attractive model of life's origins that involves as many as six polymeric types that interact to cooperatively sustain information [43]. Yet from an Occam's Razor point of view, two cooperating systems would be the most accessible to any complex system; the critical point is simply that the number must be greater than one.

In our scenario, progress towards the level of complexity characteristic of life requires the interaction of collectively autocatalytic sets (CASs). Consequently, a requisite for life is the continual functioning of the parts to sustain the whole. Life is distinct from non-life in large part because it is a Kantian Whole. There is downward selection from the whole to the parts; without the former, the latter are insignificant. Selection at every step of the process solidifies the whole (the CAS or, later, the joint CASs) and directs the chemical interactions of the parts. This is a continuously self-reinforcing phenomenon.

Importantly though, a tremendously significant *phase transition* occurs when both CASs reinforce each other. Their physical entanglement *via* the covalent bond of the mixed anhydride became manifest as coding. This may have been one of the most important phase transitions in the history of the universe in that it led to life. In the realm of particle physics, an entangled system is defined to be one whose quantum state cannot be factored as a product of states of its local constituents; that is to say, they are not individual particles but are an inseparable whole. A parallel process occurred at the macromolecular level and became life.

Previously we have noted that constraint closure drove the major transitions in the origins of life [19], and this scenario we present here is no exception. Work, the constrained release of energy into just a few degrees of freedom, was required to create the living situation. Having more than one biopolymer is itself a constraint. Another is that, in a CAS, there are more transitions than there are molecules.

Our scenario is just a working hypothesis. Yet it is amenable to possible experimental approaches for support or refutation. For example, one could put two CASs (peptides and RNAs) together and observe if their intersection leads to more than doubling in reproductive rates. A more restricted experiment would be to test if two RNA stem loops can spur the recombination or ligation of two peptides. Conversely, it should be relatively easy to test whether short peptides, through binding, help two regions of RNA that are otherwise fairly thermodynamically stable, melt.

Thinking about life and coding as mutually dependent phenomena, we may be able to address questions that could not have been answered before. Many phenomena have been discussed for 60 years regarding the code and its origins. We need to stress that life required at least two polymers. Polymers alone can perform combinatorics. Our point is that life, or Von Neuman's universal constructor, required two distinct chemistries working together; one would not be enough, and the relationship between them must have at some point been solidified as a code.

**Data accessibility.** This article has no additional data.

**Authors' contributions.** S.A.K. and N.L. devised the model. N.L. wrote the paper. Both authors read and approved the final version of the paper.

**Competing interests.** We declare we have no competing interests.

**Funding.** This project received no funding.